# Osmotic pressure: resisting or promoting DNA ejection from phage?

Meerim Jeembaeva[1], Martin Castelnovo[2], Frida Larsson[1] and Alex Evilevitch[1]*

1) *Department of Biochemistry, Center for Chemistry and Chemical Engineering, Lund University, Box 124, S-221 00, Lund, Sweden.*

2) *Laboratoire Joliot-Curie – Laboratoire de Physique, Ecole Normale Superieure de Lyon, 46 Allée d'Italie, 69364 Lyon Cedex 07, France.*

* Corresponding author: Alex.Evilevitch@biochemistry.lu.se

ABSTRACT

Recent *in vitro* experiments have shown that DNA ejection from bacteriophage can be partially stopped by surrounding osmotic pressure when ejected DNA is digested by DNase I on the course of ejection. We argue in this work by combination of experimental techniques (osmotic suppression without DNaseI monitored by UV absorbance, pulse-field electrophoresis, and cryo-EM visualization) and simple scaling modeling that intact genome (*i.e. undigested*) ejection in a crowded environment is, on the contrary, enhanced or eventually complete with the help of a pulling force resulting from DNA condensation induced by the osmotic stress itself. This demonstrates that in vivo, the osmotically stressed cell cytoplasm will promote phage DNA ejection rather than resisting it. The further addition of DNA-binding proteins under crowding conditions is shown to enhance the extent of ejection. We also found some optimal crowding conditions for which DNA content remaining in the capsid upon ejection is maximum, which correlates well with the optimal conditions of maximum DNA packaging efficiency into viral capsids observed almost 20 years ago. Biological consequences of this finding are discussed.

*Keywords: osmotic suppression, phage ejection, capsid, PEG, dextran, HU, HMGB1, DNA, phage λ.*

## 1. INTRODUCTION

Osmotic suppression experiments performed on several bacterial viruses (bacteriophages λ (Evilevitch et al., 2005; Evilevitch et al., 2003) and T5 (Castelnovo and Evilevitch, 2007)) have illustrated that phage DNA ejection in vitro is a biologically passive process (i. e., one not involving any enzymes or motors). The ejection is driven by an internal DNA "pressure" resulting from strong DNA bending and DNA-DNA repulsive forces induced by tightly packed DNA chains within the rigid phage capsid with dimensions hundreds of times smaller than the length of the phage genome (Evilevitch et al., 2008; Kindt et al., 2001; Purohit et al., 2003; Tzlil et al., 2003). Internal DNA pressures on the order of tens of atmospheres were found in these measurements, which correspond to tens of pN force. This is in agreement with the experimental force data for the inverse process of DNA-packing in λ and ϕ29 phages (Fuller et al., 2007a; Fuller et al., 2007b; Fuller et al., 2007c; Rickgauer et al., 2008; Smith et al., 2001). The osmotic suppression measurements show that DNA ejection from phage (triggered by addition of phage-receptor molecules) is partially or completely suppressed by the osmotic force of the host solution containing PEG 8000 (polyethylene glycol with MW 8000 Da) as an osmolyte. At the same time, for phage λ, we have shown that only ≈ 50% of its full-length genome (48,500 bp) will be ejected at 3 atm external osmotic pressure induced by PEG 8000 (Grayson et al., 2006), corresponding to the estimated pressure in an E. coli cell (Neidhardt, 1996) caused by highly crowded cellular cytoplasm with various macromolecules. These measurements and calculations (Evilevitch, 2006; Grayson et al., 2006; Grayson and Molineux, 2007; Inamdar et al., 2006) have raised a conceptual question – how is DNA completely internalized in crowded bacterial cytoplasm, when internal DNA pressure alone would not be sufficient for complete ejection?

Recently, we showed experimentally that non-specific DNA-binding proteins present in bacterial cytoplasm (such as HU protein) will exert significant forces on ejected phage DNA. These forces will help to "pull" the remaining DNA from capsid into cell (Evilevitch, 2006; Lof et al., 2007b). Such DNA "pulling" forces were theoretically described and attributed to "ratcheting" and "entropic Langmuir adsorption" forces acting on the DNA (Inamdar et al., 2006; Zandi et al., 2003). Similar forces can also be exerted by specific DNA-binding proteins such as RNA polymerases (Molineux, 2001). Furthermore, we showed that DNA-condensing polyamines present in bacterial cytoplasm will also exert pulling on ejected genome through condensation(Evilevitch, 2006). An alternative hydrodynamic drag model was also suggested to "flush" DNA into the cell (Grayson and Molineux, 2007). However, it was not experimentally verified, whether these forces were indeed capable of completely internalizing phage DNA against the osmotic pressure of cytoplasm. So far, this osmotic pressure caused by molecular crowding was considered as the main factor resisting phage DNA injection. As mentioned earlier, this conclusion was reached through osmotic suppression *in vitro* experiments, where DNA partially ejected from phage into PEG solution was completely digested by DNase I on the time scale of the ejection process (Evilevitch, 2006).

For the first time, we argue in this work, both by *in vitro* experiments and by physical modeling, that molecular crowding in bacterial cytoplasm, and therefore cellular osmotic pressure, *does promote* the internalization of *intact viral genome* through crowding-induced DNA condensation under *in vivo* conditions, rather than resisting its insertion. Molecular crowding usually refers to the condition when a colloidal solution, like the cytoplasm, is occupied by a significant volume of non-interacting and interacting solute molecules (Parsegian et al., 1995; Rand et al., 2000). As a rule of thumb, both type of solutes induce DNA condensation in solution, effectively acting as a poor solvent for DNA (Chebotareva et al., 2004; Hall and Minton, 2003; Kombrabail and Krishnamoorthy, 2005; Minton, 1993; Minton, 1998; Minton, 2001; Minton, 2006; Murphy and Zimmerman, 1994; Murphy and Zimmerman, 1995; Ramos et al., 2005; Sasahara et al., 2003; Stavans and Oppenheim, 2006; Vasilevskaya et al., 1995; Zimmerman and Minton, 1993; Zimmerman and Murphy, 1996). In the former case, the combination of steric competition effects and change in water activity result in water exclusion and thus "dehydration" of DNA (Rand et al., 2000). The typical example of such a phenomenon is the so-called Polymer-Salt-Induced DNA condensation (see for example: (Vasilevskaya et al., 1995). In the latter case, DNA is condensed by direct specific or non-specific interactions with DNA-binding molecules (DNA-binding proteins and polyamines present in the cell cytoplasm). When both interacting and non-interacting solutes are present in solution, the presence of the latter lowers the threshold concentration necessary to obtain *direct* condensation, resulting therefore in a synergistic condensing effect on DNA. This was demonstrated by (Murphy and Zimmerman, 1995) using a combination of both crowding and DNA-binding materials.

Molecular crowding appears to be an important factor responsible for many enzymatic processes in bacteria (Minton, 2006; Zimmerman and Murphy, 1996). It has been shown that crowding induces circularization of phage DNA (required for transcription) (Louie and Serwer, 1991; Murphy and Zimmerman, 1995; Zimmerman and Harrison, 1985), increases the rate of DNA and RNA polymerase binding to viral DNA by a factor of 100 (Sasaki et al., 2006; Zimmerman and Harrison, 1987), promotes HU-DNA binding at HU concentrations 100-fold below concentrations required for DNA condensation *in vitro* (Murphy and Zimmerman, 1995), as well as speeding up DNA digestion by DNase I (Sasaki et al., 2007). Furthermore, based on *in vitro* experiments, (Zimmerman and Murphy, 1996) have proposed that DNA is subjected to mandatory condensation in a prokaryotic cell, meaning that DNA is most of the time in a condensed state within bacteria due to molecular crowding.

In this work, we repeat the osmotic suppression measurements but with the delayed presence of DNase I in the host solution. Ejected phage DNA is then expected to be condensed as it is ejected into the crowded PEG solution before addition of DNase I. As we show below, this DNA condensation induces a *pulling force* favoring ejection, which is then capable of internalizing the entire phage genome in the cell *with the help of* its osmotic pressure. Unlike recently proposed models (Grayson and Molineux, 2007; Sao-Jose et al., 2007), we explicitly show that with increasing osmotic pressure through a given range in the host solution, ejected DNA will be more condensed resulting in stronger pulling and *not* resisting force on the phage genome. This conclusion is qualitatively supported by our theoretical modelization, which explains both the previous results of partial ejection in the presence of DNase and the new experimental results in the absence DNase. It is shown that the effect of added PEG in the solution is closely related to the induced shift in water activity. As previously described in our recent work (Evilevitch et al., 2008), an added osmolyte like PEG lowers water activity outside the capsid, resulting in a significant reduction of the DNA self-interactions inside the capsid through water escape. Therefore, the partial ejection in the presence of DNase is associated with the effects of a negative force stabilizing the stressed DNA inside the capsid. The absence of DNase in the solution induces a similar stabilizing force, or pulling force, for the ejected DNA part, due to its condensation. The subtle balance between these two forces gives rise to the behavior observed in the experiments.

This study is a systematic investigation of the DNA pulling effect on ejection from bacteriophage λ caused by condensation of the ejected genome by molecular crowding agents (PEG 8000 and dextran-12000) and DNA-binding agents (HU and HMGB1 proteins). The fraction of ejected DNA is monitored with help of cryo electron microscopy (cryo-EM), electrophoresis, and UV absorbance measurements. Based on our results, we anticipate that direct molecular crowding, together with DNA-binding molecules, will condense DNA as it is ejected in the cytoplasm of bacteria, therefore generating a strong pulling force on the phage genome. Moreover, since the capsid stays outside the cell, where osmolarity is much lower than that of cytoplasm, the stabilizing force of DNA inside the capsid is much smaller than it would be if the capsid was directly in the cytoplasm. This scenario thus strongly supports the full passive ejection of viral genome in the bacteria with the help of the osmotic pressure gradient between the cytoplasm and the extracellular medium. Therefore the present experiments, performed *with* and *without* DNase I, reconcile discrepancies between earlier observations of incomplete DNA ejection in vitro in crowded environments and complete ejection in vivo (see Figure 1).

## 2. RESULTS.

In these *in vitro* experiments, we determine the balance of forces acting on DNA inside and outside the phage λ capsid when DNA is ejected into a crowded environment. We compare two cases: when DNase I is present in the external solution and the ejected genome part is digested into nucleotides and the case without DNase when ejected DNA is condensed by the crowding agent (osmolyte) in the external solution. Combined effect of DNA-binding proteins and crowding molecules on the extent of ejection is also experimentally demonstrated. The fact that ejected phage DNA is condensed by PEG, dextran, or the DNA-binding molecules used here, at given salt and DNA concentrations, has been verified with cryo-EM of phage incubated with LamB receptor in each corresponding solution *without* DNase, as shown in Figure 2.

As is extensively described in the Discussion section, the effect of an added osmolyte like PEG in the solution of phage capsids can be fully interpreted through its effects on water activity both inside and outside the capsid. In the case of partial DNA ejection from phage, both the inside and outside parts of the genome are subjected to a *stabilizing* force which lowers the respective energetic stress on DNA, due to the decrease of water activity in the surrounding solution. For the inner part of the genome, this force tends to stop the ejection, while for the outer part of the genome, it tends to pull all the genome outside the capsid. For the sake of clarity, we term the PEG-induced forces as the *stabilizing* force for "inside" DNA and as the *pulling* force for "outside" DNA, as shown in Figure 1.

**2.1 Phage DNA ejection in the presence of molecular crowding agents (PEG or dextran).**

As mentioned in the introduction of this paper, we have used PEG 8000 and dextran-12000 to mimic the osmotic pressure and crowding of cellular cytoplasm. These two inert polymers were used in order to show that DNA condensation depends on induced osmotic stress and not on the specific identity of the polymer. In the first set of measurements, the ejection from phage was triggered by the LamB receptor in the presence of 5, 15, 20, and 30 % w/w PEG. The fraction of DNA ejected was measured using the UV-absorption method described above. This method was modified in order to determine the ejected DNA fraction *without* DNase in solution in order to allow ejected DNA to condense. Therefore, DNase was added only after the incubation of phage with LamB at different times (after the ejection process was complete, i.e. $\geq$ 30 min) to show that ejected DNA becomes condensed in PEG on the time scale of experiments where ejection and condensation processes are not time dependent at different PEG concentrations (Lof et al., 2007b; Novick and Baldeschwieler, 1988). We have also verified, using gel electrophoresis, that DNase I completely digests DNA within 30 minutes at 37°C at all the PEG concentrations used here (data not shown). As a reference sample, fraction of ejected DNA was measured *with* DNase present in PEG solution prior to LamB addition (time = 0 for DNase addition). All results are shown in Figure 3a.

At 5% w/w PEG (corresponding to an osmotic pressure of 0.5 atm) *with* DNase initially present in the sample, ejection is incomplete with 91% DNA ejected of full wt DNA length (48.5 kb). However, *without* DNase (when ejected DNA is allowed to condense first prior to DNase addition) the ejection is complete. This confirms that DNA can be completely ejected against an osmotic pressure due to the condensation of the ejected DNA fraction exerting a pulling force on the DNA. Indeed, the Cryo-EM micrograph in Figure 2a shows completely ejected phage DNA condensed by 5% w/w PEG 8000 solution.

However, at 15% w/w PEG (corresponding to 4 atm osmotic pressure), the extent of ejection is ≈ 50% *with* DNase, and only slightly higher (average value of ≈ 60%) *without* DNase. In this case, the shift in genome content stabilized within the phage is also attributed to the presence of the pulling force induced by DNA condensation.

Increasing PEG concentration to 20% w/w shifts the force balance further in favor of DNA condensed on the outside of the phage. *With* DNase, only 40% of DNA is ejected, while *without* DNase 70% of DNA is ejected. The ejection is still not complete due to the balance of two opposite forces. However, the osmotic pressure has now risen to 7 atm, while more DNA is ejected at 20% PEG than at 15% PEG *without* DNase. From this observation, we can deduce that pulling force increases faster with osmotic pressure than stabilizing force, thereby shifting the force balance towards full DNA ejection. Moreover, we observe a *maximal* DNA content (not being digested) stabilized inside the phage around 4 atm osmotic pressure, as shown in the inset plot in Figure 4. As a consequence, our data predict that this osmotic pressure condition should be optimal for the opposite process of DNA packaging. Indeed it was shown almost 20 years ago that in vitro packaging is most efficient in the presence of PEG 6000 and 8000 at concentrations corresponding to similar osmotic pressures of between 2 and 5 atm (Son et al., 1989). This observation is also consistent with the estimated *in vivo* osmotic pressure of bacterial cytoplasm of around 3 atm (Neidhardt, 1996). We anticipate therefore that these are optimal conditions for most efficient packaging of phage λ.

At 30% PEG (corresponding to 19 atm osmotic pressure) the ejection is completely suppressed *with,* and *without,* DNase, showing that stabilizing force at this osmotic pressure is high enough to overcome the capsid's stress on DNA. The pulling force for DNA outside the capsid therefore has no influence on this behavior since no DNA is ejected or condensed in the bulk solution.

Figure 3 confirms that at all PEG concentrations used here, the ejected DNA fraction does not vary with time allowed for condensation of ejected DNA prior to DNase addition (between 30 min and 4h). This shows that both ejection and condensation processes reach equilibrium within the first 30 minutes or less. In vivo, that corresponds to the situation where phage DNA is being condensed by the crowded cytoplasm as it is being injected into the cell. Condensation starts once the minimal length of DNA required for condensation has been inserted by internal pressure in the phage (Schnurr et al., 2002).

We also provide additional support for the UV-measured fraction of DNA ejected (Figure 3a) with gel electrophoresis measurements of DNA length remaining in the capsid after ejection. We incubated phage-LamB mixture in 5% w/w PEG *with* and *without* DNase. In analogy with the UV-measurements above, samples incubated in PEG *without* DNase were later mixed with DNase after 1, 2, 3, 4, and 5 hours, to separate the ejected from the unejected genome. DNA remaining in the capsid was then phenol extracted and its length was determined with gel electrophoresis (Figure 3b). In good agreement with UV measurements, *with* DNase present, only 86% of DNA was ejected (corresponding to the ~7 kb DNA band that remained in the capsid) (see second lane from left). The second, upper band in the same lane represents DNA from unopened phages (48.5 kb long), a fraction of which is always present in all samples (Evilevitch et al., 2005). At the same time, in the samples *without* DNase (lanes 3-7 from left) the ejection was complete, exhibiting only one 48.5 kb band from unopened phages. This confirms that both ejection and condensation have reached an equilibrium within the first hour or less. (Sao-Jose et al., 2007) have recently suggested that unlike electrophoresis DNA length measurements, UV measurements might underestimate the external osmotic pressure required to suppress DNA ejection (based on their electrophoresis measurements of phage SPP1 ejection). However, we have earlier compared both of these methods for suppression of DNA ejection from phage λ and did not find any discrepancies between the two data sets (Evilevitch et al., 2005). Good quantitative agreement between UV and electrophoresis data is also found in this work for all PEG concentrations. It should be noted that(Sao-Jose et al., 2007) did not perform corresponding UV measurements for phage SPP1 but instead, compared their data with earlier phage λ data (Grayson et al., 2006) collected under distinctly different salt conditions.

In order to demonstrate that condensation-induced DNA pulling depends on the crowding and osmotic stress and not on the unique properties of PEG polymer, we have repeated UV measurements of ejected DNA fraction from λ using dextran as a crowding agent instead of PEG. We chose dextran-12000 (~11,600 MW) since it has molecular dimensions (Stoke's radius) similar to that of PEG 8000 (Kuga, 1981) and therefore should provide similar crowding effect on DNA at equivalent osmotic pressures. The buffer conditions were otherwise identical to solutions with PEG (50 mM Tris-HCl and 10 mM $MgSO_4$, pH 7.4). We chose to incubate phage with LamB in 11 and 22% w/w dextran-12000 (corresponding to 0.5 atm and 2 atm osmotic pressures, respectively (Ogston and Wells, 1970)) in order to test whether the shifts in ejected length observed with PEG at similar osmotic pressures could be reproduced with dextran. We have determined ejected DNA fraction from phage at these dextran concentrations *with* and *without* DNase with help of the UV measurements described above. In order to quantify ejected DNA fraction in the latter case, DNase I was added to the samples 1 hour after incubation with LamB at 37°C, since we have shown above that both ejection and condensation have reached equilibrium after that time (Figure 4). This figure shows that 85% of DNA is ejected in 11% w/w dextran *with* DNase initially present, in good agreement with corresponding 5% w/w PEG 8000 data (with 91% DNA ejected) at the same osmotic pressure of 0.5 atm. Also in similarity to the PEG case *without* DNase, when DNase is added only 1 hour after incubation, the ejection is complete, demonstrating the pulling effect of dextran-induced DNA condensation. The cryo-EM micrograph in Figure 2b shows a phage-LamB solution incubated in 11% w/w dextran-12000 with ejected DNA condensed by dextran. Thus, it is not a specific PEG property, but crowding and osmotic stress effects of the polymer that condense the ejected DNA and induce a net pulling force which results in a complete ejection against an osmotic pressure.

In 22% w/w dextran corresponding to 2 atm, *with* DNase present, 64% DNA was ejected. The ejected fraction corresponded well to the value obtained earlier with 10% w/w PEG 8000 at the same osmotic pressure (Grayson et al., 2006). However, *without* DNase (when DNase was added 1h after incubation with LamB), the ejection was still not complete, with 81 % DNA ejected, confirming the behavior observed with PEG at similar osmotic pressures. More precisely, 22% dextran provided a slightly lower osmotic pressure (2atm) compared to 15% PEG (4 atm) for which 60% DNA was ejected *without* DNase.

**2.2 Phage DNA ejection in the presence of both molecular crowding and DNA-binding agents (PEG and HU/HMGB1).**

Thus, as illustrated above, DNA condensation induced by crowding and the resulting pressure increase in osmotic pressure is sufficient to pull DNA out from the phage against an osmotic pressure induced by the crowding agents themselves. We propose that a similar pulling effect will be in fact achieved by essentially any DNA condensation mechanism. In bacterial cytoplasm, cellular DNA is compacted by several factors. The most important of these are: macromolecular crowding (Murphy and Zimmerman, 1995); DNA-binding proteins (such as HU and H-NS proteins) (Drlica and Rouviere-Yaniv, 1987; Sarkar et al., 2007); and polyamines (Tabor and Tabor, 1985). The cryo-EM images in Figure 2 show ejected phage λ DNA that has been condensed by PEG 8000 (Figure 2a), dextran-12000 (Figure 2b), spermine (Figure 2c), HU (Figure 2d),

HMGB1 (Figure 2e) and E. coli cytoplasm extract (Figure 2f). In particularly, as already mentioned, there is an indirect DNA condensation mechanism through a synergistic effect of crowding that significantly enhances binding of DNA-binding proteins to DNA (Murphy and Zimmerman, 1995). Therefore we expect that if DNA ejection is triggered in the presence of both crowding and DNA-binding agents not entering the capsid, the partial ejection should shift further toward full ejection, since DNA-binding agents will increase the net pulling force.

In order to test this hypothesis, we determined the length of unejected λ-DNA remaining in the capsid in the presence of both HU protein and PEG 8000 using pulse-field gel electrophoresis. We choose 15% w/w PEG 8000 as a reference solution (corresponding to the osmotic pressure of 4 atm found in bacteria). Under these reference conditions, UV measurements showed that only 60% of the DNA is ejected from the phage *without DNase* present (Figure 3a). The results of the pulse-field electrophoresis determination of the unejected DNA length are shown in Figure 5a. The leftmost lane, labeled ''ladder,'' is an 8 − 48.5 kb standard consisting of 13 bands of DNA of known length. The first lane to the right of the ladder is the band corresponding to the DNA extracted from an unopened λ-capsid, which is 48.5 kb for wt phage λ. The next lane to the right corresponds to DNA extracted when LamB is added to the phage solution in 15% PEG *with* DNase present. Here we see the result of a partial ejection that left a DNA length of ~19.4 - 29.9 kb inside, and also a second band of 48.5 kb from the fraction of the phage that remains unopened (this fraction is observed in all samples (Evilevitch et al., 2005)). We see a large variation in the unejected DNA length in 15 % PEG compared to PEG concentrations below and above this value (see 5%, 10 and 20% PEG data shown

in(Evilevitch et al., 2005). Average DNA length is 24.7 kb or 49% of ejected genome in agreement with UV data shown in Figure 3a. In the next lane, phage was incubated with LamB in 15% PEG for 2 hours prior to DNase addition to allow the ejected DNA to condense. DNA length remaining in the capsid corresponded to an even broader band distribution of DNA lengths of ~11 - 30 kb. This corresponds to an average of 20 kb, or ~60% ejected λ–genome, confirming therefore the result of the UV measurements. The next lane shows the same solution conditions as in the previous lane (*without* DNase) but with HU protein added to the solution (with final HU dimer concentration of 475 nM, approximately 100 times lower than cellular HU concentrations (Azam et al., 1999)). This time, only ~10 – 20 kb DNA remained in the capsid after ejection is complete, corresponding to an average of 15 kb or ~70% ejected genome. Thus, we clearly observe an increased pulling on phage genome due to combined condensation effect of HU and PEG. However, the ejection is still not complete since both stabilizing and pulling forces are present *in vitro,* unlike the *in vivo* case. In the latter case, the stabilizing force component is expected to be much smaller since phage remains outside the cell during its genome ejection into E. coli cell. The DNA-condensation pulling force induced by crowding and DNA-binding agents might then be more than sufficient to complete the ejection. We also confirmed with gel electrophoresis (data not shown) that HU binding to DNA does not prevent DNase from completely digesting DNA in the presence of PEG within the time allowed for DNA digestion in all experiments.

The observed combined effect of molecular crowding and DNA-binding agent on the extent of DNA ejection from capsids might also be relevant in the case of eukaryotic cell infection. Although during infection of an eukaryotic cell, the whole viral capsid enters the cell, it is still not fully understood how the genome of dsDNA viruses is released from the capsid, especially in those cases where the capsid does not disassemble during infection (as in the case of Herpes Simplex Virus (HSV-1) (Newcomb et al., 2007; Ojala et al., 2000; Shahin et al., 2006)). As in the case of motor packaged dsDNA viruses like Herpes, DNA can exert internal capsid pressure that can promote DNA ejection into cell nucleolus. However, this pressure alone would not be sufficient to complete ejection due to osmotic pressure in the cell, as has been argued for phage ejection. Unlike phage, the eukaryotic virus would be exposed to change in water activity surrounding the viral genome both inside and outside the capsid. This case scenario would thus correspond to our experimental *in vitro* system with phage, where DNA is affected by PEG both inside and outside the capsid. Eukaryotic cytoplasm exhibits just as crowded an environment as bacterial cytoplasm, with a large number of DNA-binding and compacting molecules (Ovadi and Saks, 2004). Presumably, a similar DNA-pulling mechanism can complete viral DNA release through condensation of ejected DNA induced by crowding combined with DNA-binding agents in the cell (Salman et al., 2001). In mammalian cells, HMGB1 protein has been shown to be functionally similar to HU in bacteria. It is a non-specifically DNA-binding protein, also present at micromolar concentrations (McCauley et al., 2007; Megraw and Chae, 1993; Skoko et al., 2004). Figure 2e shows ejected phage λ DNA (after incubation with LamB) condensed by 475 nM of rat HMGB1 without PEG present. With the above-proposed motivation, we have also investigated the effect of indirect crowding (induced by PEG 8000) in combination with rat HMGB1 on the extent of phage DNA ejection. We have incubated phage λ for 2 hours at 37°C with LamB in 15% PEG 8000 and 475 nM HMGB1. In that case, DNase I was added in order to separate the ejected DNA part from the unejected, and after phenol extraction, the purified DNA was loaded on a pulse-field gel (Figure 5a). The right-most lane on the gel shows that no DNA was left in the phage despite the force stabilizing DNA in the capsid and thus resisting ejection induced by 4 atm pressure of 15% PEG 8000 (only 48.5 kb DNA band from unopened phage fraction is observed). Thus, HMGB1 promotes viral DNA pulling to an even greater degree than HU protein under crowded conditions, presumably due to stronger compaction of DNA (given that both proteins were present at the same concentration). This also demonstrates, that any DNA condensation, even that caused by bacterially unrelated DNA-condensing proteins, will induce DNA pulling and complete viral ejection against an osmotic pressure (in this case as high as 4 atm corresponding to in vivo pressure). Figures 5b and 5c show cryo-EM micrographs of two samples: a phage-LamB solution incubated in 15% w/w PEG 8000 *without* (Figure 5b), and one *with* HMGB1 (Figure 5c). Figure 5b shows partially DNA-filled phage particles and condensed DNA aggregates formed from DNA ejected from several phage particles. It should be noted here that very few such DNA condensates were observed in comparison to the 5% w/w PEG sample). Figure 5c shows *empty* phage particles co-aggregated with their ejected and condensed DNA, confirming the pulse-field electrophoresis results. It can also be noted that when the ejected λ-DNA is condensed in PEG or in HMGB1 alone (Figures 2a and 2e), phage particles are not co-aggregated with the DNA condensate as they are in Figure 5c with both condensing factors present. This also illustrates an indirect crowding effect induced by DNA-binding proteins in the crowding media, as described in (Murphy and Zimmerman, 1995). The cryo-EM micrographs shown in Figures 2 and 5 are typical representations from tens of images collected for each sample.

## 3. DISCUSSION.

**3.1 New theoretical interpretation of osmotic suppression experiments.**

The results of the experiments presented above are readily understood using the simple model presented below, based on the interpretation of osmotic phenomena using solvent activity modulations. In particular, we use a simple scaling description of DNA energetics inside and outside the capsid that allows us to identify the balance of the most relevant contributions. A detailed version of the present theory, as well as an exhaustive and quantitative comparison with experimental data is beyond the scope of this work, and will be presented elsewhere (M.C., A.E. manuscript in preparation). Rather, we introduce here the alternative solvent-based interpretation of the osmotic suppression experiments, which is qualitatively consistent both with previous and present experimental data. The advantage of the scaling description is that the focus can be placed on the relevant behavior of the system, regardless of precise microscopic details.

The interpretation of osmotic suppression experiments is deeply associated with the balance of solvent molecules inside and outside the capsid, as we recently suggested in another work (Evilevitch et al., 2008). It is precisely this balance that gives rise to an osmotic pressure difference between the two regions. However, increasing the concentration of added osmolyte in the bathing solution leads to a decrease of solvent activity, or chemical potential, outside the capsid. As a consequence, solvent molecules tend to escape from the unfavorable environment of the region inside the capsid containing a given amount of packaged DNA. This effect lowers DNA-DNA repulsive interactions inside the capsid through changes in the hydration properties and therefore contributes to the stabilization of stressed DNA inside the capsid. This leads to an alternative interpretation of osmotic suppression experiments when DNase I is present in solution: the ejection force at equilibrium acting on the partially ejected DNA is balanced by the stabilizing force inside the capsid induced by the presence of the outer osmolytes. A natural consequence of the present interpretation is that osmotic suppression experiments in the presence of DNase provide a direct experimental evidence of the communication between non-contacting macromolecules (Volker and Breslauer, 2005). This is a concept whose importance has been slowly emerging over recent years in biological systems as is shown by the increasing usage of osmotic stress methods (Parsegian et al., 2000). When no DNase is present in the solution, the part of ejected DNA in the outer solution is condensed in the presence of the added osmolytes. From the point of view of solvent molecules, this situation is very similar to the one previously described, i.e.increasing the concentration of added osmolytes lowers the solvent activity outside the volume of the DNA coil, therefore inducing a significant lowering of self-interactions among DNA monomers through solvent escape from the coil region. Above some critical concentration value for osmolytes, the nucleic acid eventually condenses into a toroid. In the particular geometry of DNA translocating from inside of the capsid towards the outside medium through the phage's tail, this condensation generates a force pulling on the DNA remaining in the capsid. The magnitude of this force is proportional to the difference between the chemical potential the DNA monomers in the unperturbed coil and the condensed conformation. Based on the present solvent-based approach, we conclude that added osmolytes have two major influences on intact DNA ejection from bacteriophage when no DNase is present in solution: *(i)* they reduce the ejection force associated with the part of the DNA present in the capsid through an osmotic stabilizing force; and *(ii)* they increase the force pulling the DNA outside the capsid. It is the imbalance of these two forces that gives rise to incomplete ejection behavior observed in our experiments, as discussed below. In subsequent sections, we first calculate the scaling behavior of the force pulling the DNA outside the capsid, then the scaling behavior of the force stabilizing DNA inside the capsid. The reference state for all free energy calculations associated with different DNA conformations is taken as rod-like straight DNA.

**3.2 Scaling behavior of pulling force outside the capsid.**

The derivation of the pulling force is based on earlier work by (Odijk, 1998b) and (Tzlil et al., 2003). While the former work gave the general outline for an analytical computation of toroid characteristics using mainly electrostatic short-range DNA self-interactions, the latter used, rather, a phenomenological description of these interactions in hexagonal phases based on osmotic stress measurements. The resulting scaling of DNA toroid energetics is the same, and will be used here. The leading order energetic balance describing the density and shape of a DNA toroidal condensate is composed of three terms: bulk attractive interactions proportional to the total volume of the condensate (and therefore to the total length of DNA), surface energy associated with the finite size of the condensate and the bending energy associated with the toroidal conformation. The last two terms scale similarly for the size of toroid that minimizes the total free energy.

The free energy of the outer DNA condensate is therefore written as:

$$\frac{G_{outside}(L, P_{osm})}{kT} = -\epsilon(P_{osm})L + \gamma(P_{osm})L^{3/5}$$

where $L$ is the DNA length outside the capsid, $\epsilon(P_{osm})$ is the bulk energy per unit length gained by condensation, and $\gamma(P_{osm})$ is a parameter that represents the balance between surface tension and bending energy. For the sake of clarity, this term will be referred to as the surface tension term. The precise dependence of these two parameters on osmotic pressure is model-dependent. At the scaling level, it is sufficient to know that both parameters increase with osmotic pressure. The derivation of this free energy with respect to DNA length gives the force pulling on the nucleic acid:

$$F_{outside}(L, P_{osm}) = -kT\epsilon(P_{osm}) + \frac{3kT\gamma(P_{osm})}{5}L^{-2/5}$$

This force is negative, favoring the growth of the toroid, and it increases slightly as more DNA is ejected. At this step, it should be noted that this negative force takes implicitly into account the work involved in the insertion of the toroid inside the solution. As a consequence, the osmotic pressure does not *resist* but rather *promotes* DNA ejection. This is further demonstrated theoretically elsewhere (MC., AE. manuscript in preparation).

**3.3 Scaling behavior of the stabilizing force inside the capsid.**

As it was already mentioned at the beginning of the Discussion Section, the added osmolytes in the solution containing the capsids induce a similar lowering of self-interactions between DNA monomers inside the capsid through solvent escape from the DNA condensate volume (which might differ from the full volume of the capsid, due to the geometrical constraints associated with DNA packaging in such a small volume (Odijk, 1998b) for example). One should therefore expect a similar scaling dependence for the stabilizing force inside the capsid. The major difference comes from the strong confinement or constraint imposed by the rigid walls of the capsid. This is accounted for at the scaling level by two modifications of the toroid free energy. The first modification is the presence of a higher order term $w_{ev}L^{\alpha}$, with $\alpha > 1$ taking into account the excluded volume interactions for confinement of the toroid and where $w_{ev}$ is a constant whose interpretation is similar to a virial coefficient. The precise value of the exponent $\alpha$ is not important since it has been verified *a posteriori* that it leads to the same scaling behavior of the system and therefore the same qualitative conclusions (MC, AE manuscript in preparation). The second modification is a shift in the surface tension term $(\gamma(P_{osm}) + \gamma_{cap})L^{3/5}$, taking into account, in a generic way, the modification of surface tension properties of DNA condensate inside the capsid. Indeed, direct interaction between the DNA condensate and the capsid wall might induce such a change. It turns out *a posteriori* that the case of a negative $\gamma_{cap}$, and therefore a net decrease of surface tension as compared to the outside condensate, is indeed able to explain the peculiar behavior of the equilibrium length remaining inside the capsid at osmotic pressures higher than 0.5 atm (e.g. corresponding to PEG concentration 15 and 20% w/w). The total free energy of the inside-DNA condensate can therefore be written as:

$$\frac{G_{inside}(L, P_{osm})}{kT} = -\epsilon(P_{osm})L + (\gamma(P_{osm}) + \gamma_{cap})L^{3/5} + w_{ev}L^{\alpha}$$

This scaling model of inner DNA condensate captures the most relevant energetic balance factors, although it omits to describe in a precise way the conformation of packaged DNA inside the capsid.

In order to interpret the partial ejection observed in the presence of DNase, which digests outer DNA into single nucleotides on the time course of ejection (Lof et al., 2007a), one needs simply to consider the free energy of inner DNA as described by the previous equation, because the reference state used for its calculation is straight DNA, similar to the part of DNA remaining undigested outside the capsid (smaller than 150 bp, which is the typical length scale of straight conformation) This free energy has an osmotic-pressure-dependent minimum resulting, in general, from the balance of bulk attractive energy ($\sim -\epsilon(P_{osm})L$) and repulsive confinement energy ($\sim +w_{ev}L^{\alpha}$). The values of equilibrium DNA length remaining inside as measured by osmotic suppression experiments are therefore given by this osmotic pressure-dependent minimum, as typically shown in Figure 6, in the curve labeled "in vitro + DNase" with the arrow at $L_1$ indicating the free energy minimum.

**3.4 Balance of forces: in vitro and in vivo case.**

The sum of the two free energies described above allows us to analyze the equilibrium ejection behavior when no DNase is present in solution, therefore allowing outer DNA to undergo toroidal condensation. The total free energy for partially ejected DNA of full length $L_{tot}$ reads:

$$G_{tot}(L, P_{osm}) = G_{inside}(L, P_{osm}) + G_{outside}(L_{tot} - L, P_{osm})$$

The addition of a free energy favoring the growth of outer toroid leads necessarily to a shift of the previous balance towards more ejected DNA, in qualitative accordance with experimental measurements. This behavior is clearly understood as one analyzes separately the role of different energetic contributions for the inner condensate. When both surface tension changes associated with the capsid and confinement energy are neglected in a first approximation, the two ideal (unconstrained) toroids are in equilibrium. In this academic case, there are only two possible equilibrium states, corresponding to two free energy minima: all the DNA remains inside or outside. In other words, when two undeformed toroids are in equilibrium, the largest toroid will always unwind the smallest. When confinement energy is taken into account, but still without surface tension changes due to the capsid, the free energy has two minima, corresponding to full ejection or partial ejection with more than half of the genome content still inside the capsid. The lowest minima correspond to full ejection. This behavior has been recently demonstrated by numerical simulations investigating the influence of poor solvent conditions on the extent of polymer ejection from a non-interacting capsid (Ali et al., 2008). It was concluded that under poor solvent conditions, ejection is always complete. This has to be contrasted with our experimental data, which show partial ejection of intact DNA at 2 atm (22% dextran), 4 atm (15% PEG) and at 7 atm (20% PEG). The amount of DNA remaining inside the capsid at all these osmotic pressures is below 50% of whole wt genome (see inset in Figure 4).

We found that the stabilization of DNA in the presence of the pulling force induced by outer DNA condensation is accounted for by the presence of a negative shift in surface tension. The presence of this extra term shifts the full ejection free energy minimum mentioned above toward the finite length remaining in the capsid. This is illustrated in Figure 6 in the curve labeled "in vitro - DNase", with the arrow at $L_2$ indicating the free energy minimum. For the sake of clarity, the minimum is highlighted at a different scale in the inset of Figure 6. The present scaling model therefore supports the presence of a pulling force associated with DNA condensation. In our *in vitro* experiments, where DNA both inside and outside the capsid is influenced by the same modulation of solvent activity through added osmolytes, this leads to partial ejection behavior (except at lower osmotic pressures, such as 0.5 atm, where the ejection is nevertheless complete). The model can now be extrapolated to predict the consequence of the *in vivo* ejection process. It is known that in this case, the virion binds to its receptor (LamB in the case of phage λ) on the bacterial surface. DNA is then injected into the cytoplasm while the capsid stays outside in the extracellular space (ES). Different water activities are expected inside and outside the bacteria due to the strong molecular crowding in the cytoplasm. This situation can therefore be simulated using our scaling model with different osmotic pressure for the DNA inside the capsid (in the extracellular space, *ES*, environment with osmotic pressure $P_{ES}$), and DNA outside the capsid (in the cytoplasmic environment with osmotic pressure $P_{cyto} > P_{ES}$). The free energy corresponding to the *in vivo* ejection process is therefore written as:

$$G_{in\,vivo}(L, P_{cyto}, P_{ES}) = G_{inside}(L, P_{ES}) + G_{outside}(L_{tot} - L, P_{cyto})$$

As an example, the curve associated with this free energy is shown in Figure 6 with the label "in vivo - DNase" for $P_{cyto} = 4 atm$ and $P_{ES} = 0.5 atm$, and compared to the *in vitro* case with the same osmotic pressure of 4 atm for DNA inside and outside the capsid. These values are thought to be representative of osmotic pressure *in vivo*. The choice of the latter low osmotic pressure value in the extra cellular space is qualitatively justified by the presence of salts and nutrients. While the *"in vitro"* free energy exhibits partial ejection behavior as shown by the presence of a finite length minimum, the *in vivo* free energy exhibits full ejection (as indicated by the arrow *L₃* in Figure 6). This is to be expected since in the *in vivo* situation, the stabilizing force inside the capsid is much weaker than the force pulling DNA inside the cytoplasm. As a consequence, in addition to our experimental observations, the scaling model also provides strong support for full *in vivo* internalization of the λ genome with the help of cytoplasmic osmotic pressure. It should also be noted that we tested a broad range of parameters $\epsilon(P_{osm}), \gamma(P_{osm}), \gamma_{cap}, w_{ev}$ together with some variation of the exponents of the surface tension terms (inside and outside the capsid), and the confinement term. It turns out that the conclusions drawn from the choice of the present parameters are rather robust and general in this respect. Moreover, a more precise modeling of both inner and outer condensate beyond the scaling level, along the line of previous modeling by (Odijk, 1998a), (Tzlil et al., 2003), and (Purohit et al., 2005) show the same behavior (MC, AE manuscript in preparation).

## 4. SUMMARY AND CONCLUSIONS.

In this work, we have presented the results of osmotic suppression experiments of intact genomes from phage λ, using a new protocol that does not involve immediate DNase I addition, instead of the previous one where ejected DNA was digested by DNase I concurrently with the ejection. In addition, we tested the influence of DNA-binding proteins like bacterial HU and mammalian HMGB1. The results were complemented by pulse-field electrophoresis, cryo-EM visualization, and scaling modeling. Our main conclusion is that DNA condensation induced either by neutral osmolytes, by DNA-binding proteins, or by the synergetic action of both, exerts a pulling force that shifts the partial ejection behavior towards a larger measure of ejection and eventually towards full ejection under certain conditions.

In the absence of DNA-binding proteins, both DNA remaining in the capsid and that which is ejected are submersed into osmotically-stressed solution of PEG or dextran.

This introduces two opposite forces on the DNA: one stabilizing DNA inside phage capsid, thus resisting ejection and the other condensing the ejected DNA part and pulling it out from the phage. We show that these two forces can be balanced in vitro, with incomplete ejection as a result. Nevertheless, even then, we show that at lower osmotic pressures (such as 0.5 atm set by 5% PEG 8000 or 11% dextran-12000), the DNA-pulling force dominates and is capable of pulling DNA completely from the phage despite the presence of force-stabilizing DNA inside the capsid induced by the osmolyte. We show also that at higher osmotic pressures ($\geq 2$ atm), a larger DNA fraction is ejected when DNA is allowed to condense, again due to the presence of the pulling force. Even then, the ejection is complete if DNA condensation is further enhanced by the combination of DNA-binding protein and a crowding agent such as HMGB1 and 15% w/w PEG 8000 (at 4 atm). The presence of both condensation factors also significantly decreases the amount of each factor required for condensation. These experimental results can be qualitatively described by a scaling model of DNA condensation and confinement. The quantitative comparison between theory and experiments requires more precise modeling and is beyond the scope of this work (MC., AE. manuscript in preparation).

As a consequence, in the *in vivo* situation, we expect the force-stabilizing DNA inside the phage to be strongly decreased since the phage particle stays outside the cell, meaning that the encapsidated DNA is exposed to a much lower osmotic pressure than in the cytoplasm. On the other hand, the pulling force on the ejected DNA increases with increasing osmotic pressure due to enhanced DNA compaction. *Since our experimental results confirm the presence of a strong pulling force induced by DNA condensation in the crowded cytoplasmic environment, absence of a DNA stabilizing force in vivo will* *ultimately result in complete internalization of the phage genome, against an osmotic pressure gradient.* We also see evidence of this, using our scaling model, which shows that the pulling force resulting from DNA condensation by bacterial cytoplasm is sufficient to completely internalize DNA at sufficient osmotic pressure in the cell. As we have shown *in vitro*, several force factors by themselves can condense DNA. *In vivo,* these factors are combined, thus providing an "excess" of direct and indirect crowding forces over what is needed for condensation. We conclude that the osmotic pressure and molecular crowding in general are promoting DNA internalization into the cell, rather than resisting it.

Moreover, our experimental results show that there is a maximal amount of DNA that remains inside the capsid once ejection is triggered by the receptor at an osmotic pressure of roughly 4 atm (corresponding to estimated bacterial osmotic pressure). This implies that the reverse process of DNA packaging into bacteriophage should be most efficient under these conditions, since it is associated with the optimal balance between the stabilizing force helping the packaging, and the pulling force resisting the packaging. This is consistent with previous observations made in the group of P. Serwer (Son et al., 1989), showing that there exists optimal osmolyte conditions (corresponding to osmotic pressures between 2 and 5 atm) for which DNA packaging into bacteriophage is the most efficient.

## 5. MATERIALS AND METHODS

**5.1 Bacteriophage strain and preparation of phage stock.**

Wild type (wt) bacteriophage λ with genome length 48.5 kbp was produced by thermal induction of lysogenic *Escherichia coli* strain AE1 derived from strain S2773. Phage purification details are described elsewhere (Evilevitch et al., 2003). Phage was purified by CsCl equilibrium centrifugation. The sample was dialyzed from CsCl against TM buffer (10 mM $MgSO_4$ and 50 mM Tris-HCl/pH 7.4). The final titer was $10^{13}$ virions/ml, determined by plaque assay.

**5.2 Preparation of LamB phage λ receptor.**

Phage λ receptor was the LamB protein purified from pop 154, a strain of *E. coli* K12 in which the *lamB* gene has been transduced from *Shigella sonnei* 3070. This protein has been shown to cause complete in vitro ejection of DNA from λ within seconds at 37°C, in the absence of the added solvents required with the wild-type E. coli receptor. Purified LamB was solubilized from the outer membrane with the detergent, octyl polyoxyethylene (octyl-POE).

**5.3 UV measurements of osmotic suppression.**

The method for quantifying the amount of DNA ejected has been described in (Evilevitch et al., 2003). Phage λ and the LamB receptor were incubated in TM buffer containing 5, 15, 20, and 30% w/w PEG 8000 (corresponding to osmotic pressures of 0.5, 4, 7, and 19 atm) at 37 $^0$C. The phage : receptor ratio was 1 virion per 1000 LamB trimers. In order to allow phage to eject DNA without DNase I present and only after the ejection, to quantify the fraction of ejected genome, DNase I was added to the reaction mixture only after 0.5, 1, 2, 3 and 4 hours and the sample was additionally incubated for 1 hour to allow complete digestion of ejected DNA. As a reference sample to quantify fraction of DNA ejected in PEG with DNase I present, DNase I was added to the phage solution prior to LamB addition. All samples were then centrifuged at 115 000 x g in a type A-100 aluminum fixed-angle rotor in a Beckman airfuge for 2 hours to pellet the capsids at $14^0$C. Ejected and digested DNA nucleotides remaining in the supernatant were quantified by UV absorbance at 260 nm (Agilent 8453 UV-visible Spectroscopy System). The ejected DNA fraction was determined from [Abs(phage + PEG + DNase I + LamB) – Abs(phage + PEG + DNase I)] divided by [Abs(phage + LamB + DNase I) – Abs(phage + DNase I)], where (phage + PEG + DNase I) and (phage + DNase I) correspond to the two negative controls and (phage + LamB + DNase I) to the one positive control (see details in (Evilevitch et al., 2003)).

These measurements were also repeated with dextran-12000 (MW~11600 Da) obtained from SigmaAldrich as an osmolite, instead of PEG 8000. Final dextran concentrations were 11 and 22 % w/w (corresponding to 0.5 atm and 2 atm osmotic pressures, respectively).

**5.4 Cryo transmission electron microscopy (cryo-EM).**

Phage λ incubated with LamB in various media specimens for electron microscopy was prepared in a controlled environment vitrification system (CEVS) to ensure fixed temperature and to avoid water losses from the solution during sample preparation. The specimens were prepared as thin liquid films, < 0.3 mm thick, on hydrophilic (glow discharged) lacey carbon films, supported by a copper grid and quenched into liquid ethane at its freezing point. The technique was described in detail in (Bellare et al., 1988). The technique leads to vitrified specimens, so that component segmentation, rearrangement and water crystallization are prevented and original microstructures are preserved during thermal fixation. The vitrified specimens were stored under liquid nitrogen and transferred to the electron microscope (Philips CM 120 BioTWIN) equipped with a post-column energy filter, using an Oxford CT3500 cryo-holder and its workstation. The accelerating voltage was 120 kV with a magnification of 45,000x and a nominal defocus of 1 μm. The images were recorded digitally with a CCD camera (Gatan MSC791).

**5.5 DNA extraction and electrophoresis.**

DNA extraction from phage with phenol is described in(Evilevitch et al., 2005). Extracted DNA was resolved on an electrophoretic gel with 1% agarose (Top Vision LE GQ Agarose) in 1xTAE buffer (40mM Tris, 20mM acetic acid, 1mM EDTA) to resolve DNA fragments below ~20 kb. Voltage for agarose gel was 7 V/cm for 45 minutes. A GeneRuler DNA ladder from Fermentas (composed of 15 chromatography-purified individual DNA fragments 20,000 - 75 bp) and full length wt λ-DNA (48.5 kb) were used as markers. Samples were pre-heated at 65 $^0$C for 5 min and cooled on ice before loading them on the gel to avoid closing of the cohesive ends of the λ-DNA. Gels were stained with SYBR gold dye for 15 minutes and washed in MQ water.

**5.6 Pulse-field gel electrophoresis.**

We used a 1% pulse-field certified agarose gel in 0.5x TBE buffer (1x TBE solution: 89 mM Tris, 89 mM boric acid, 2 mM EDTA, pH-8.4) at 14$^0$C for 24.3 hours at angle 120 $^0$C in CHEF MAPPER XA (BioRad). Switch time was ramped algorithmically from 0.22 sec to 0.92 seconds with ramp factor 0.357, forward voltage 9V/cm and reverse voltage was 6 V/cm. CHEF DNA marker (8-48 kb) from Biorad and λ-DNA (Invitrogen) were used as ladders.

**ACKNOWLEDGEMENTS**

We would like to acknowledge Alexei Khokhlov for valuable discussions. Mark Williams and Remus Thei Dame for providing HMGB1 and HU proteins used in this study. This work was supported through grants from the Swedish Research Council and Crafoord Foundation (to AE).

**Figure Captions**

**Figure 1:** Illustration of DNA ejection in the crowded environment of PEG molecules *with* and *without* DNase I. In the case of partial DNA ejection from phage *with* DNase I, no DNA condensate is formed outside. Thus, only the inside part of the genome is subjected to a *stabilizing* force lowering the respective energetic stress on the DNA, due to the decrease of water activity in the surrounding solution. This force stops the ejection. However, *without* DNase I, both the inside and outside parts of the genome are subjected to this *stabilizing* force. For the inner part of the genome, this force tends to stop the ejection, while for the outer part of the genome, it tends to pull all the genome outside the capsid. We term the PEG induced forces as the *stabilizing* force for "inside" DNA and *pulling* force for "outside" DNA. In vivo, the *pulling* force is higher than the *stabilizing* force due to the osmotic pressure difference. This results in a complete ejection.

**Figure 2:** Cryo-EM images of phage λ incubated with LamB receptor at 37°C in: (A) 5% w/w PEG 8000, (B) 11% w/w dextran-12000, (C) 1 mM spermine, (D) 475 nM HU protein, (E) 475 nM HMGB1 protein and (F) E. coli lysate. Figure shows ejected phage λ DNA that has been condensed by each of these solutions, respectively.

**Figure 3:** (A) The fraction of DNA ejected from phage λ in the presence of 5, 15, 20 and 30 % w/w PEG 8000, measured with the UV-absorption. DNase I was added at 0.5, 1, 2, 3 and 4 hours after the incubation with LamB at 37°C to allow condensation of ejected DNA. After that, all samples were incubated for 1 hour with DNase I in order to completely digest ejected DNA fraction measured here. As a reference, the fraction of ejected DNA was also determined *with* DNase present in the PEG solution prior to LamB addition (time = 0 for DNase addition). Dashed lines are drawn to guide the eye. (B) Complementary gel electrophoresis measurement of DNA length remaining in the capsid after ejection in 5% w/w PEG 8000 *with* and *without* DNase. Samples were first incubated in PEG *without* DNase. DNase was later added after 1, 2, 3, 4 and 5 hours to separate ejected from unejected genome. A reference sample *with* DNase initially present corresponds to t = 0.

**Figure 4:** UV-measured DNA ejected fraction from phage λ incubated with LamB in 11 and 22% w/w dextran-12000 (corresponding to 0.5 atm and 2 atm osmotic pressures). Ejected DNA fraction is determined *with* and *without* DNase I. In the latter case, DNase I is added to the samples 1 hour after incubation with LamB at 37°C (time sufficient to allow ejected DNA to condense in the presence of dextran). The inset shows ejected DNA fraction *without* DNase I versus external osmotic pressure, ranging from 0 to 7 atm (combined from measurements with PEG and dextran). The data shows a minimum in the ejected DNA fraction corresponding to a *maximal* DNA content stabilized inside the phage at around 4 atm osmotic pressure. This osmotic pressure condition should be optimal for the opposite process of DNA packaging.

**Figure 5:** (A) Pulse-field gel electrophoresis measurement of length remaining in the capsid after DNA ejection from phage λ incubated with LamB at 37°C *without* DNase I in 15% w/w PEG 8000 with 475 nM of HU and HMGB1 proteins added (two separate

samples). After 2 hours incubation, DNase I was added in order to digest ejected and condensed DNA and thus separate ejected from unejected genome parts. Cryo-EM micrographs of: (B) phage-LamB solution incubated in 15% w/w PEG 8000 *without* (C) and *with* HMGB1 protein. Both samples are *without* DNase I added. Figure 5 (B) shows partially DNA-filled phage particles, while Figure 5 (C) shows *empty* phage particles co-aggregated with their ejected and condensed DNA, in agreement with pulse-field electrophoresis results.

**Figure 6:** Free energy of DNA partially ejected as a function of DNA length remaining inside. The parameters of the scaling model are $\epsilon(P_{osm})$=0.5 $P_{osm}$ (pN), $\gamma(P_{osm})$=6.96 $P_{osm}^{(2/3)}$ (pN.nm$^{(2/5)}$)), $\gamma_{cap}$=-4 (pN.nm$^{(2/5)}$), $w_{ev}$=2.88 10$^{-17}$ (pN.nm$^{-4}$). In these equations, osmotic pressure is expressed in *atm,* and the scaling exponent of surface tension with the osmotic pressure comes from the scaling properties of polymer solutions "in vitro +DNase": free energy inside at 4 atm osmotic pressure; "in vitro – DNase": sum of free energies inside and outside at 4 atm osmotic pressure; "in vivo –DNase": free energy inside at 4 atm and outside at 0.5atm. The three arrows indicate the locations of minima of each free energy. The inset shows a magnification of the "in vitro-DNase curve.

# REFERENCES:

Ali, I., D. Marenduzzo, andJ.M. Yeomans. 2008. Ejection dynamics of polymeric chains from viral capsids: effect of solvent quality. *Biophys J.*

Azam, T.A., A. Iwata, A. Nishimura, S. Ueda, andA. Ishihama. 1999. Growth phase-dependent variation in protein composition of the Escherichia coli nucleoid. *Journal of Bacteriology* 181(20):6361-6370.

Bellare, J.R., H.T. Davis, L.E. Scriven, andY. Talmon. 1988. Controlled environment vitrification system: an improved sample preparation technique. *J Electron Microsc Tech* 10(1):87-111.

Castelnovo, M., andA. Evilevitch. 2007. DNA ejection from bacteriophage: Towards a general behavior for osmotic-suppression experiments. *Eur Phys J E Soft Matter* 24(1):9-18.

Chebotareva, N.A., B.I. Kurganov, andN.B. Livanova. 2004. Biochemical effects of molecular crowding. *Biochemistry-Moscow* 69(11):1239-+.

Drlica, K., andJ. Rouviere-Yaniv. 1987. Histonelike proteins of bacteria. *Microbiol Rev* 51(3):301-319.

Evilevitch, A. 2006. Effects of condensing agent and nuclease on the extent of ejection from phage lambda. *J Phys Chem B Condens Matter Mater Surf Interfaces Biophys* 110(44):22261-22265.

Evilevitch, A., L.T. Fang, A.M. Yoffe, M. Castelnovo, D.C. Rau, V.A. Parsegian, W.M. Gelbart, andC.M. Knobler. 2008. Effects of salt concentrations and bending energy on the extent of ejection of phage genomes. *Biophys J* 94(3):1110-1120.

Evilevitch, A., J.W. Gober, M. Phillips, C.K. Knobler, andW.M. Gelbart. 2005. Measurements of DNA lengths remaining in a viral capsid after osmotically suppressed partial ejection. *Biophysical Journal* 88(1):751-756.

Evilevitch, A., L. Lavelle, C.M. Knobler, E. Raspaud, andW.M. Gelbart. 2003. Osmotic pressure inhibition of DNA ejection from phage. *Proc. Nat. Acad. Sci. USA* 100(16):9292–9295.

Fuller, D.N., D.M. Raymer, V.I. Kottadiel, V.B. Rao, andD.E. Smith. 2007a. Single phage T4 DNA packaging motors exhibit large force generation, high velocity, and dynamic variability. *Proc Natl Acad Sci U S A* 104(43):16868-16873.

Fuller, D.N., D.M. Raymer, J.P. Rickgauer, R.M. Robertson, C.E. Catalano, D.L. Anderson, S. Grimes, andD.E. Smith. 2007b. Measurements of single DNA molecule packaging dynamics in bacteriophage lambda reveal high forces, high motor processivity, and capsid transformations. *J Mol Biol* 373(5):1113-1122.

Fuller, D.N., J.P. Rickgauer, P.J. Jardine, S. Grimes, D.L. Anderson, andD.E. Smith. 2007c. Ionic effects on viral DNA packaging and portal motor function in bacteriophage phi 29. *Proc Natl Acad Sci U S A* 104(27):11245-11250.

Grayson, P., A. Evilevitch, M.M. Inamdar, P.K. Purohit, W.M. Gelbart, C.M. Knobler, andR. Phillips. 2006. The effect of genome length on ejection forces in bacteriophage lambda. *Virology* 348(2):430-436.

Grayson, P., andI.J. Molineux. 2007. Is phage DNA 'injected' into cells--biologists and physicists can agree. *Curr Opin Microbiol* 10(4):401-409.

Hall, D., and A.P. Minton. 2003. Macromolecular crowding: qualitative and semiquantitative successes, quantitative challenges. *Biochim Biophys Acta* 1649(2):127-139.

Inamdar, M.M., W.M. Gelbart, and R. Phillips. 2006. Dynamics of DNA ejection from bacteriophage. *Biophys J* 91(2):411-420.

Kindt, J., S. Tzlil, A. Ben-Shaul, and W.M. Gelbart. 2001. DNA packaging and ejection forces in bacteriophage. *Proceedings of the National Academy of Sciences of the United States of America* 98(24):13671-13674.

Kombrabail, M.H., and G. Krishnamoorthy. 2005. Fluorescence dynamics of DNA condensed by the molecular crowding agent poly(ethylene glycol). *J Fluoresc* 15(5):741-747.

Kuga, S. 1981. Pore-Size Distribution Analysis of Gel Substances by Size Exclusion Chromatography. *Journal of Chromatography* 206(3):449-461.

Lof, D., K. Schillen, B. Jonsson, and A. Evilevitch. 2007a. Dynamic and static light scattering analysis of DNA ejection from the phage lambda. *Phys Rev E Stat Nonlin Soft Matter Phys* 76(1 Pt 1):011914.

Lof, D., K. Schillen, B. Jonsson, and A. Evilevitch. 2007b. Forces controlling the rate of DNA ejection from phage lambda. *J Mol Biol* 368(1):55-65.

Louie, D., and P. Serwer. 1991. Effects of temperature on excluded volume-promoted cyclization and concatemerization of cohesive-ended DNA longer than 0.04 Mb. *Nucleic Acids Res* 19(11):3047-3054.

McCauley, M.J., J. Zimmerman, L.J. Maher, 3rd, and M.C. Williams. 2007. HMGB binding to DNA: single and double box motifs. *J Mol Biol* 374(4):993-1004.

Megraw, T.L., and C.B. Chae. 1993. Functional complementarity between the HMG1-like yeast mitochondrial histone HM and the bacterial histone-like protein HU. *J Biol Chem* 268(17):12758-12763.

Minton, A.P. 1993. Macromolecular crowding and molecular recognition. *J Mol Recognit* 6(4):211-214.

Minton, A.P. 1998. Molecular crowding: analysis of effects of high concentrations of inert cosolutes on biochemical equilibria and rates in terms of volume exclusion. *Methods Enzymol* 295:127-149.

Minton, A.P. 2001. The influence of macromolecular crowding and macromolecular confinement on biochemical reactions in physiological media. *J Biol Chem* 276(14):10577-10580.

Minton, A.P. 2006. Macromolecular crowding. *Curr Biol* 16(8):R269-271.

Molineux, I.J. 2001. No syringes please, ejection of phage T7 DNA from the virion is enzyme driven. *Mol Microbiol* 40(1):1-8.

Murphy, L.D., and S.B. Zimmerman. 1994. Macromolecular crowding effects on the interaction of DNA with Escherichia coli DNA-binding proteins: a model for bacterial nucleoid stabilization. *Biochim Biophys Acta* 1219(2):277-284.

Murphy, L.D., and S.B. Zimmerman. 1995. Condensation and cohesion of lambda DNA in cell extracts and other media: implications for the structure and function of DNA in prokaryotes. *Biophys Chem* 57(1):71-92.

Neidhardt, F., editor. 1996. Escherichia Coli and Salmonella Typhimurium. ASM Press.

Newcomb, W.W., F.P. Booy, and J.C. Brown. 2007. Uncoating the herpes simplex virus genome. *J Mol Biol* 370(4):633-642.

Novick, S.L., and J.D. Baldeschwieler. 1988. Fluorescence Measurement of the Kinetics of DNA Injection by Bacteriophage-Lambda into Liposomes. *Biochemistry* 27(20):7919-7924.

Odijk, T. 1998a. Hexagonally packed DNA within bacteriophage T7 stabilized by curvature stress. *Biophys J* 75(3):1223-1227.

Odijk, T. 1998b. Osmotic compaction of supercoiled DNA into a bacterial nucleoid. *Biophys Chem* 73(1-2):23-29.

Ogston, A.G., and J.D. Wells. 1970. Osmometry with single sephadex beads. *Biochem J* 119(1):67-73.

Ojala, P.M., B. Sodeik, M.W. Ebersold, U. Kutay, and A. Helenius. 2000. Herpes simplex virus type 1 entry into host cells: reconstitution of capsid binding and uncoating at the nuclear pore complex in vitro. *Mol Cell Biol* 20(13):4922-4931.

Ovadi, J., and V. Saks. 2004. On the origin of intracellular compartmentation and organized metabolic systems. *Mol Cell Biochem* 256-257(1-2):5-12.

Parsegian, V.A., R.P. Rand, and D.C. Rau. 1995. Macromolecules and water: probing with osmotic stress. *Methods Enzymol* 259:43-94.

Parsegian, V.A., R.P. Rand, and D.C. Rau. 2000. Osmotic stress, crowding, preferential hydration, and binding: A comparison of perspectives. *Proc Natl Acad Sci U S A* 97(8):3987-3992.

Purohit, P.K., M.M. Inamdar, P.D. Grayson, T.M. Squires, J. Kondev, and R. Phillips. 2005. Forces during bacteriophage DNA packaging and ejection. *Biophys J* 88(2):851-866.

Purohit, P.K., J. Kondev, and R. Phillips. 2003. Mechanics of DNA packaging in viruses. *Proc. Nat. Acad. Sci. USA* 100(6):3173-3178.

Ramos, J.E., Jr., R. de Vries, and J. Ruggiero Neto. 2005. DNA psi-condensation and reentrant decondensation: effect of the PEG degree of polymerization. *J Phys Chem B* 109(49):23661-23665.

Rand, R.P., V.A. Parsegian, and D.C. Rau. 2000. Intracellular osmotic action. *Cell Mol Life Sci* 57(7):1018-1032.

Rickgauer, J.P., D.N. Fuller, S. Grimes, P.J. Jardine, D.L. Anderson, and D.E. Smith. 2008. Portal motor velocity and internal force resisting viral DNA packaging in bacteriophage phi29. *Biophys J* 94(1):159-167.

Salman, H., D. Zbaida, Y. Rabin, D. Chatenay, and M. Elbaum. 2001. Kinetics and mechanism of DNA uptake into the cell nucleus. *Proc Natl Acad Sci U S A* 98(13):7247-7252.

Sao-Jose, C., M. de Frutos, E. Raspaud, M.A. Santos, and P. Tavares. 2007. Pressure built by DNA packing inside virions: enough to drive DNA ejection in vitro, largely insufficient for delivery into the bacterial cytoplasm. *J Mol Biol* 374(2):346-355.

Sarkar, T., I. Vitoc, I. Mukerji, and N.V. Hud. 2007. Bacterial protein HU dictates the morphology of DNA condensates produced by crowding agents and polyamines. *Nucleic Acids Res* 35(3):951-961.

Sasahara, K., P. McPhie, and A.P. Minton. 2003. Effect of dextran on protein stability and conformation attributed to macromolecular crowding. *J Mol Biol* 326(4):1227-1237.

Sasaki, Y., D. Miyoshi, and N. Sugimoto. 2006. Effect of molecular crowding on DNA polymerase activity. *Biotechnol J* 1(4):440-446.

Sasaki, Y., D. Miyoshi, and N. Sugimoto. 2007. Regulation of DNA nucleases by molecular crowding. *Nucleic Acids Res* 35(12):4086-4093.

Schnurr, B., F. Gittes, and F.C. MacKintosh. 2002. Metastable intermediates in the condensation of semiflexible polymers. *Physical Review E* 65(6):-.

Shahin, V., W. Hafezi, H. Oberleithner, Y. Ludwig, B. Windoffer, H. Schillers, and J.E. Kuhn. 2006. The genome of HSV-1 translocates through the nuclear pore as a condensed rod-like structure. *J Cell Sci* 119(Pt 1):23-30.

Skoko, D., B. Wong, R.C. Johnson, and J.F. Marko. 2004. Micromechanical analysis of the binding of DNA-bending proteins HMGB1, NHP6A, and HU reveals their ability to form highly stable DNA-protein complexes. *Biochemistry* 43(43):13867-13874.

Smith, D.E., S.J. Tans, S.B. Smith, S. Grimes, D.L. Anderson, and C. Bustamante. 2001. The bacteriophage phi 29 portal motor can package DNA against a large internal force. *Nature* 413(6857):748-752.

Son, M., S.J. Hayes, and P. Serwer. 1989. Optimization of the in vitro packaging efficiency of bacteriophage T7 DNA: effects of neutral polymers. *Gene* 82(2):321-325.

Stavans, J., and A. Oppenheim. 2006. DNA-protein interactions and bacterial chromosome architecture. *Phys Biol* 3(4):R1-10.

Tabor, C.W., and H. Tabor. 1985. Polyamines in microorganisms. *Microbiol Rev* 49(1):81-99.

Tzlil, S., J.T. Kindt, W.M. Gelbart, and A. Ben-Shaul. 2003. Forces and pressures in DNA packaging and release from viral capsids. *Biophys. J.* 84:1616-1627.

Vasilevskaya, V.V., A.R. Khokhlov, Y. Matsuzawa, and K. Yoshikawa. 1995. Collapse of Single DNA Molecule in Poly(Ethylene Glycol) Solutions. *Journal of Chemical Physics* 102(16):6595-6602.

Volker, J., and K.J. Breslauer. 2005. Communication between noncontacting macromolecules. *Annu Rev Biophys Biomol Struct* 34:21-42.

Zandi, R., D. Reguera, J. Rudnick, and W.M. Gelbart. 2003. What drives the translocation of stiff chains? *Proc Natl Acad Sci U S A* 100(15):8649-8653.

Zimmerman, S.B., and B. Harrison. 1985. Macromolecular crowding accelerates the cohesion of DNA fragments with complementary termini. *Nucleic Acids Res* 13(7):2241-2249.

Zimmerman, S.B., and B. Harrison. 1987. Macromolecular crowding increases binding of DNA polymerase to DNA: an adaptive effect. *Proc Natl Acad Sci U S A* 84(7):1871-1875.

Zimmerman, S.B., and A.P. Minton. 1993. Macromolecular crowding: biochemical, biophysical, and physiological consequences. *Annu Rev Biophys Biomol Struct* 22:27-65.

Zimmerman, S.B., and L.D. Murphy. 1996. Macromolecular crowding and the mandatory condensation of DNA in bacteria. *FEBS Lett* 390(3):245-248.

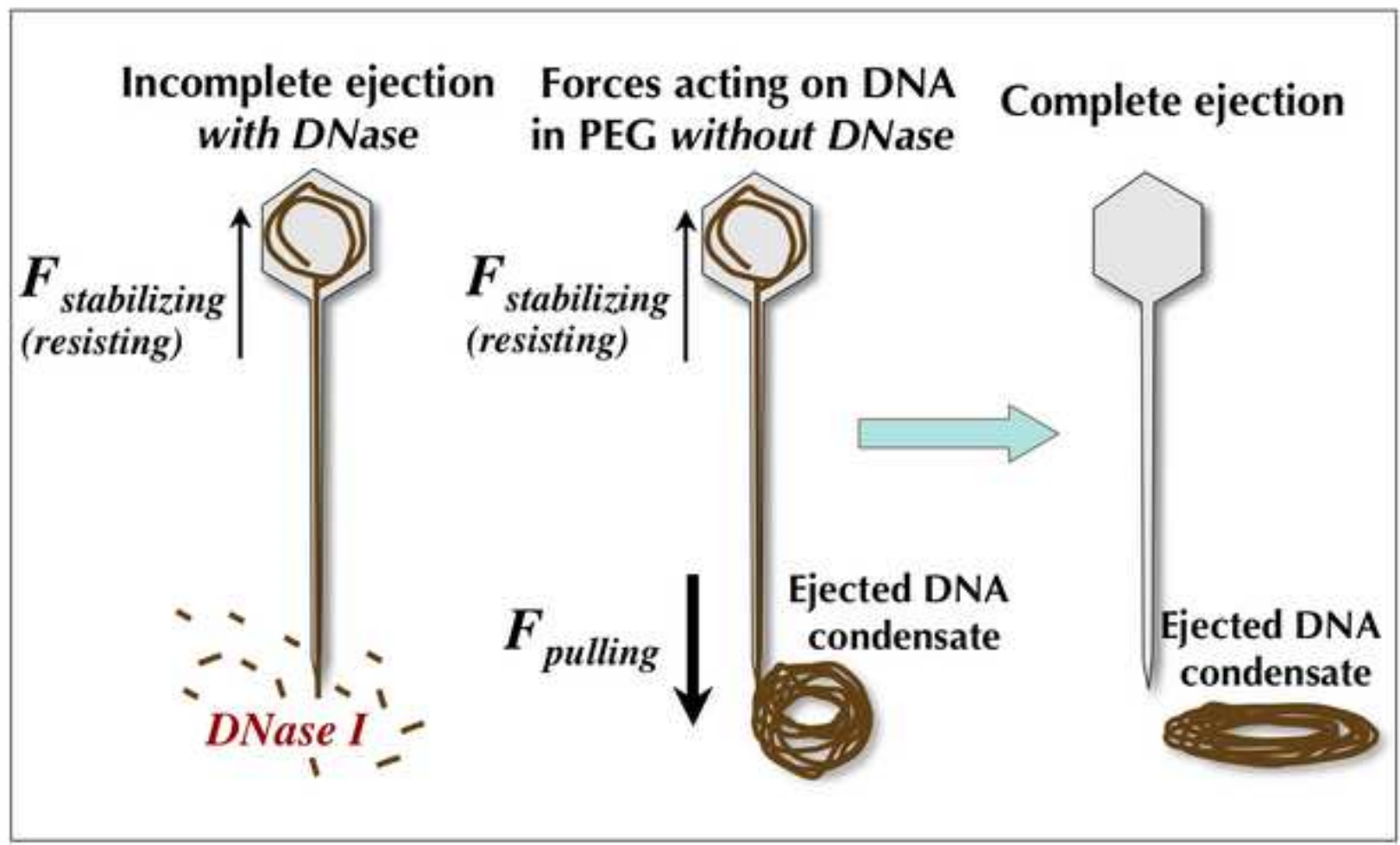

**Figure 2**
[Click here to download high resolution image]

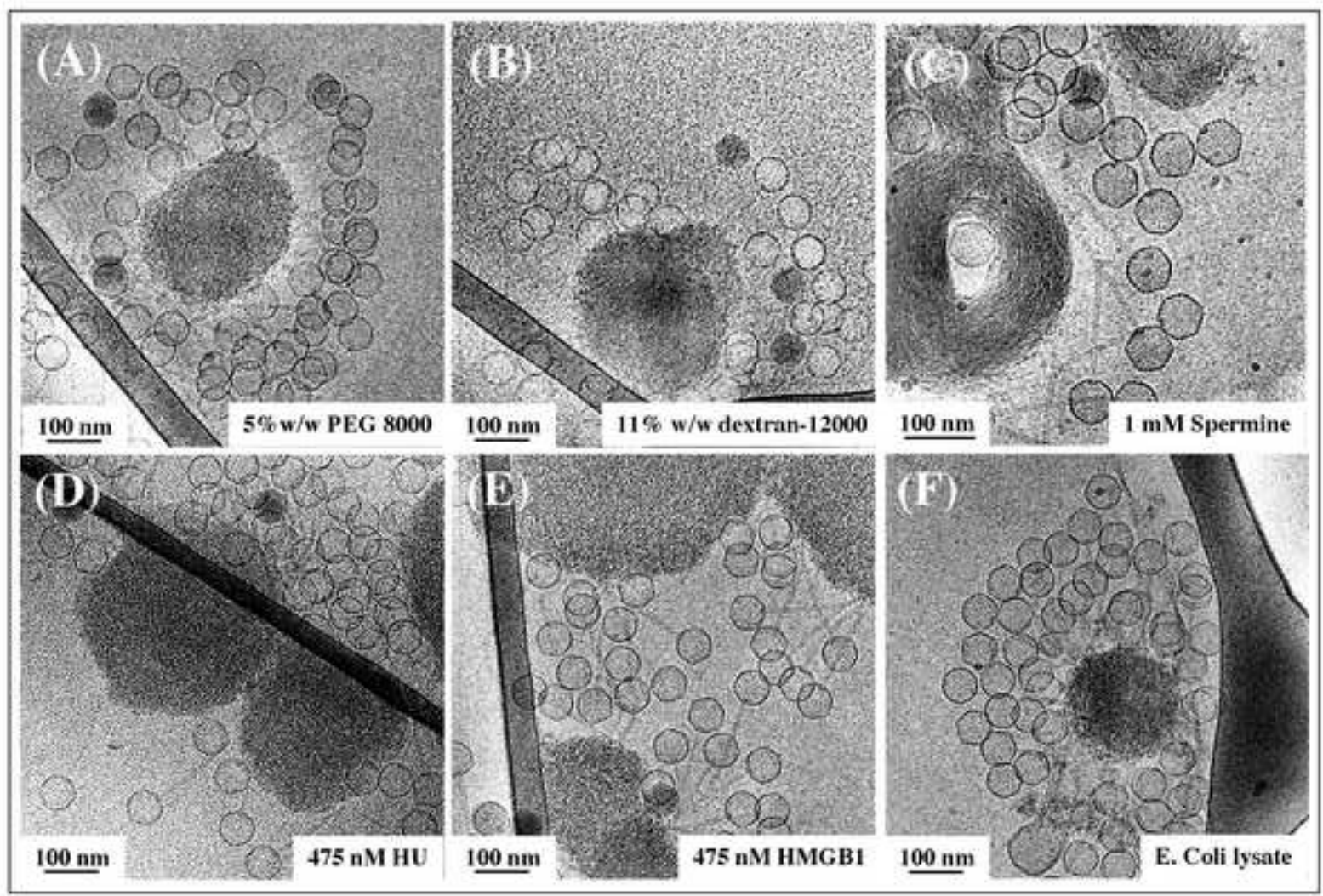

**Figure 3**
**Click here to download high resolution image**

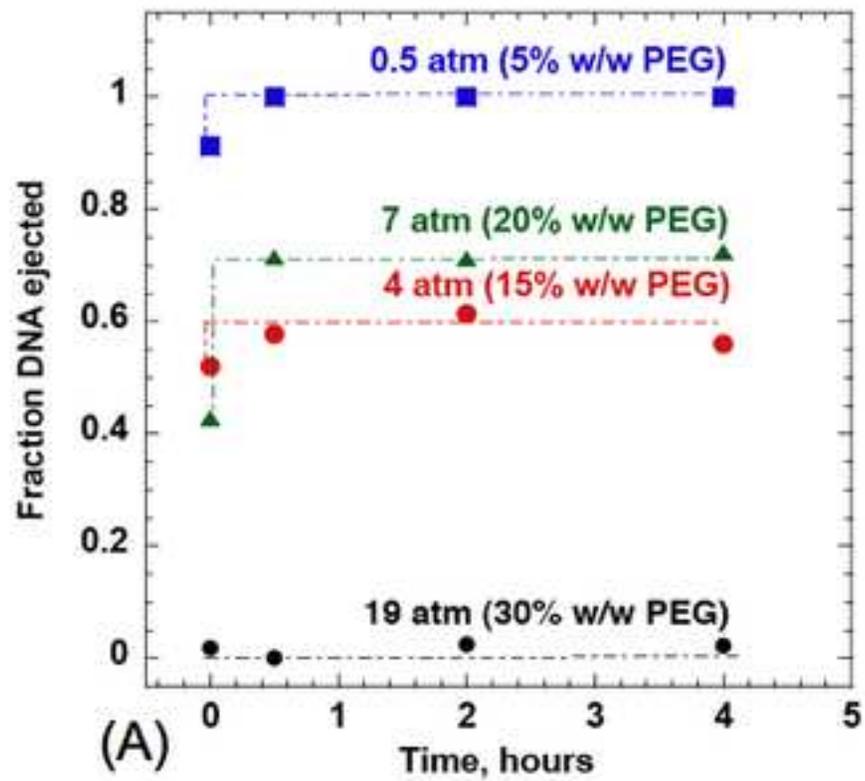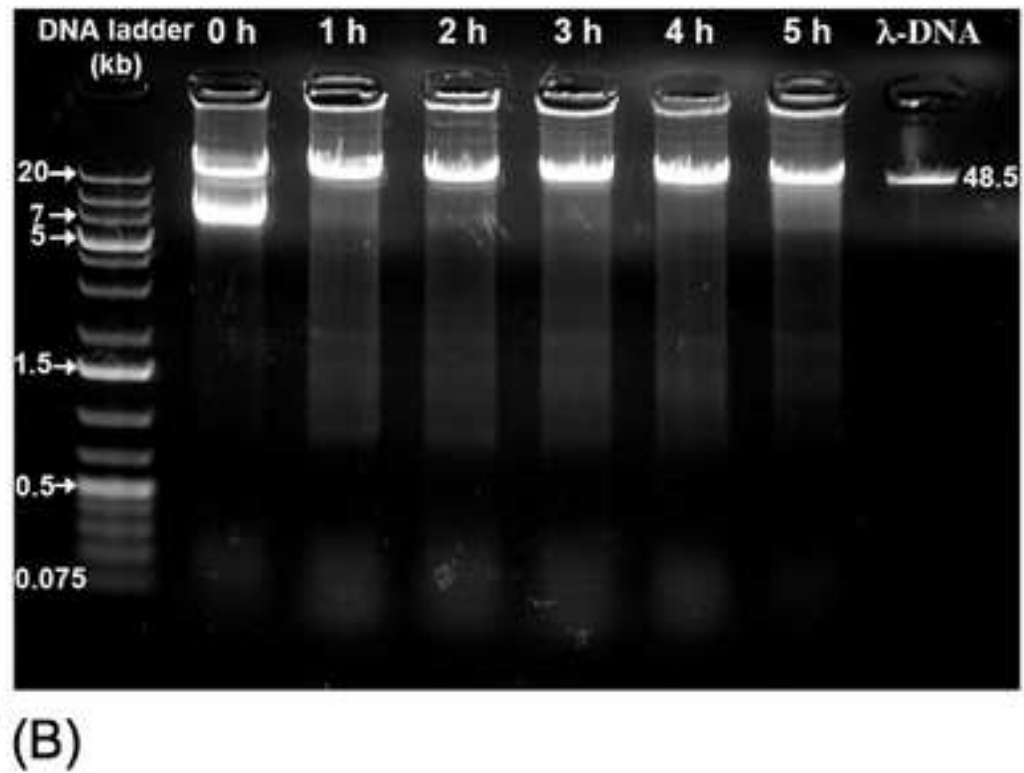

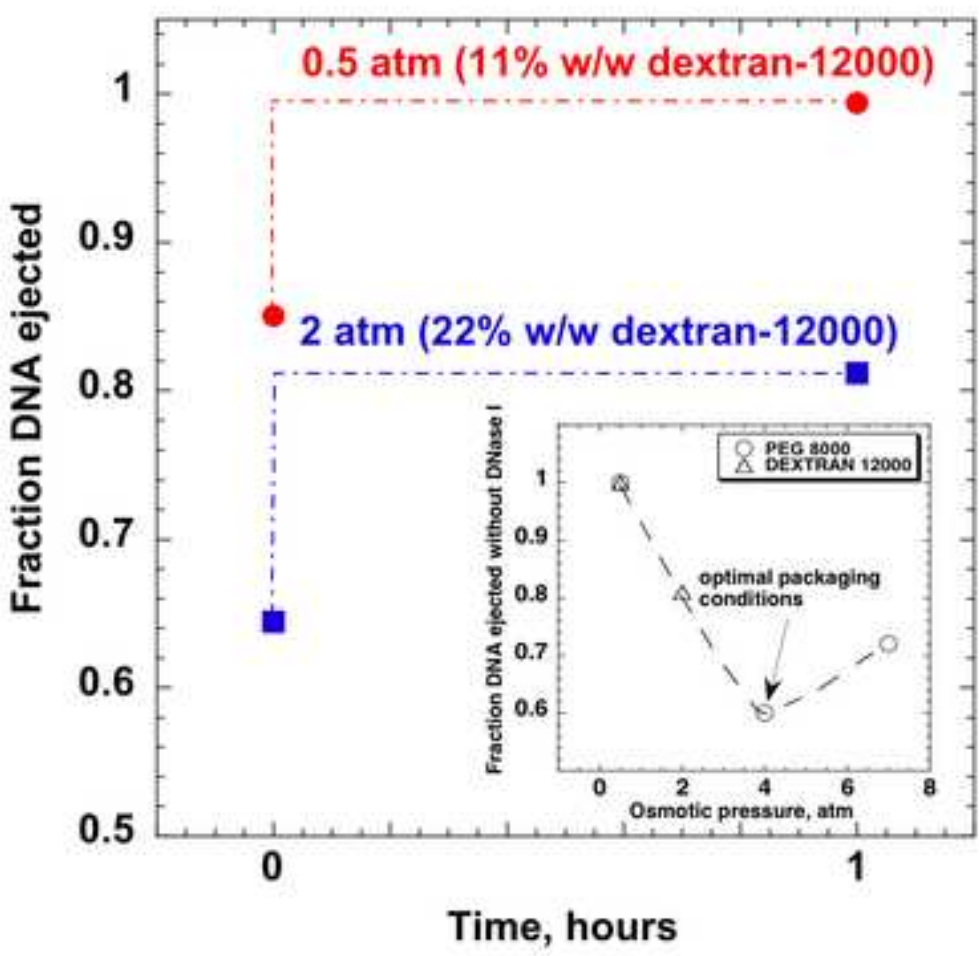

**Figure 5**
**Click here to download high resolution image**

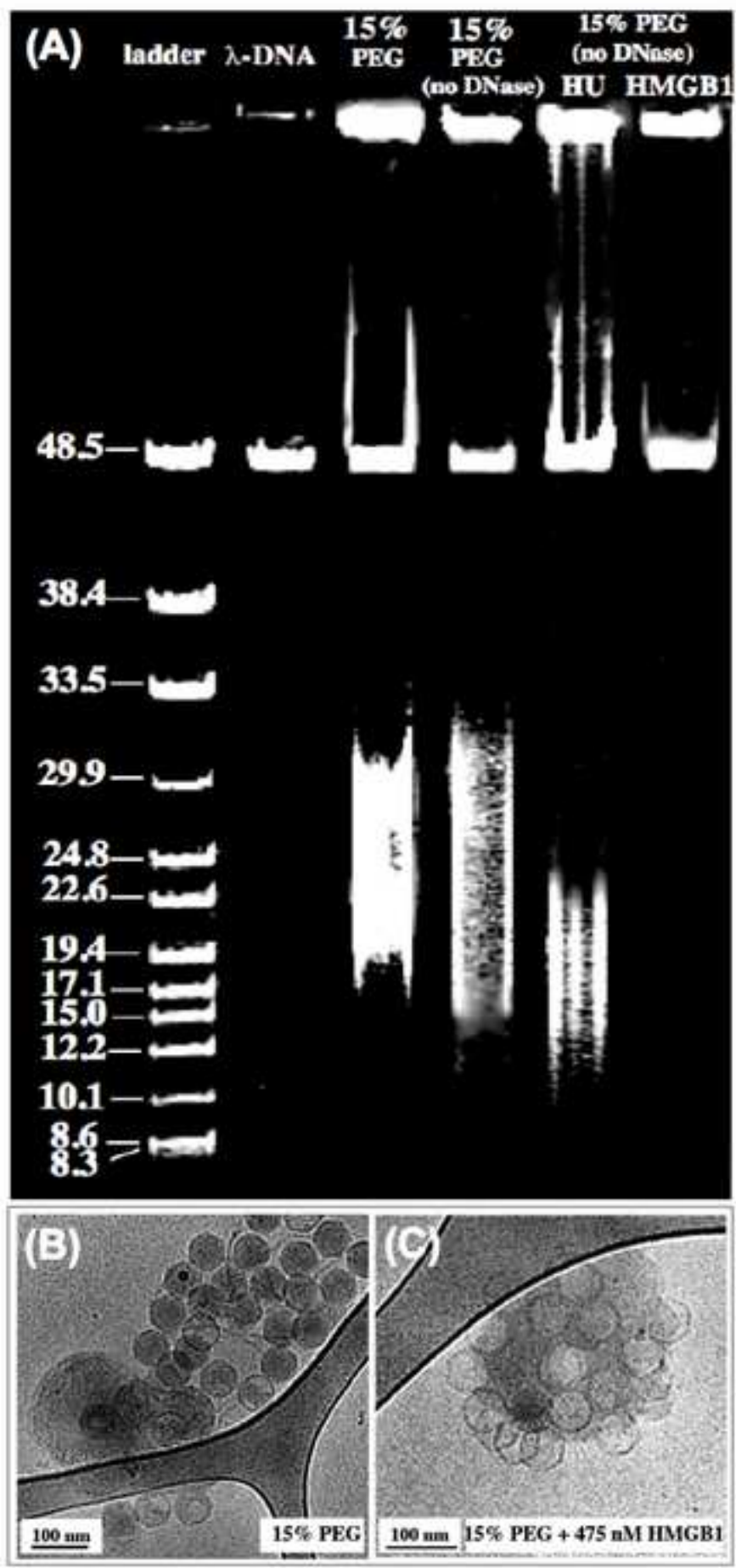

**Figure 6**
**Click here to download high resolution image**

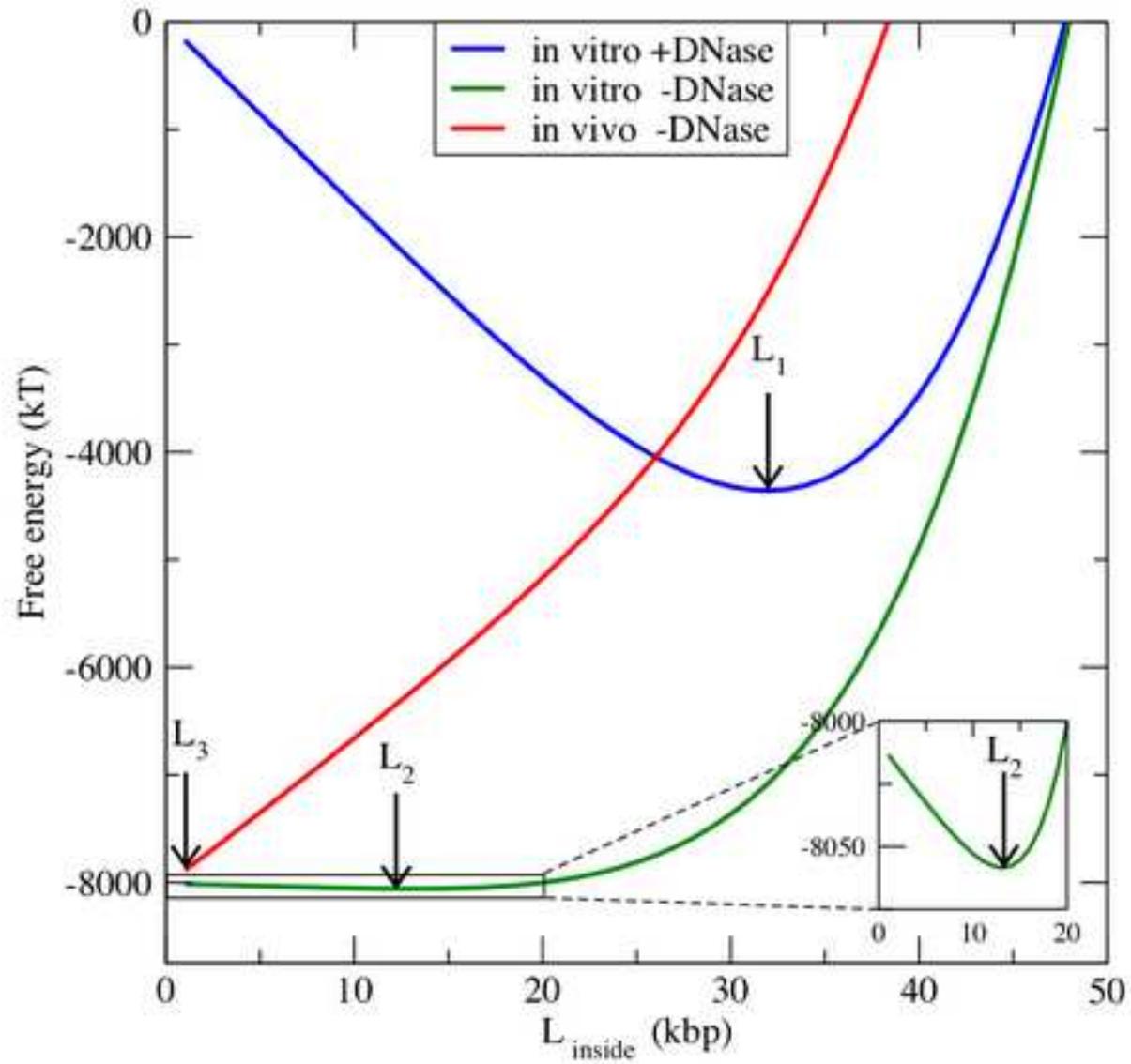